\documentclass[traditabstract]{aa}

\usepackage{graphicx}
\usepackage{txfonts}
\usepackage{natbib}
\usepackage{amssymb}
\usepackage{amsfonts}
\usepackage{amsbsy}
\usepackage{amsmath}
\usepackage{pdflscape}
\newcommand{\PFA}{P_{\rm FA}}
\newcommand{\PD}{P_{\rm D}}

\newcommand{\Cb}{\boldsymbol{C}}
\newcommand{\fb}{\boldsymbol{f}}
\newcommand{\gb}{\boldsymbol{g}}

\newcommand{\Hb}{\boldsymbol{H}}
\newcommand{\nb}{\boldsymbol{n}}
\newcommand{\ssb}{\boldsymbol{s}}

\newcommand{\wb}{\boldsymbol{w}}
\newcommand{\xb}{\boldsymbol{x}}

\newcommand{\Hc}{\mathcal{H}}
\newcommand{\Tc}{\mathcal{T}}

\newcommand{\Fmatb}{\boldsymbol{{\mathcal F}}}
\newcommand{\Nmatb}{\boldsymbol{{\mathcal N}}}
\newcommand{\Smatb}{\boldsymbol{{\mathcal S}}}
\newcommand{\Xmatb}{\boldsymbol{{\mathcal X}}}

\begin{document}

\title{On the Correct Estimate of the Probability of False Detection \\ of the Matched Filter  in Weak-Signal Detection Problems}

   \author{R. Vio
          \inst{1}
          \and
          P. Andreani 
         \inst{2}
         }

   \institute{Chip Computers Consulting s.r.l., Viale Don L.~Sturzo 82,
              S.Liberale di Marcon, 30020 Venice, Italy\\
              \email{robertovio@tin.it}
         \and
           ESO, Karl Schwarzschild strasse 2, 85748 Garching, Germany \\
             \email{pandrean@eso.org}
             }

   \date{Received....; accepted....}

   \abstract{The detection reliability of weak signals is a critical issue in many astronomical contexts and may have severe consequences for determining number counts and luminosity functions, but also for optimizing the use of telescope time in follow-up observations. Because of its optimal properties, one of the most popular and widely-used detection technique is the matched filter (MF). This is a linear filter designed to maximise the detectability of a signal of known structure that is buried in additive Gaussian random noise. In this work we show that in the very common situation where the number and position of the searched signals within a data sequence (e.g. an emission line in a spectrum) or an image (e.g. a point-source in an interferometric map) are unknown, this technique, when applied in its standard form, may severely underestimate the probability of false detection. This is because the correct use of the MF relies upon a priori knowledge of the position of the signal of interest. In the absence of this information, the statistical significance of features that are actually noise is overestimated and detections claimed that are actually spurious. For this reason, we present an alternative method of computing the probability of false detection that is based on the probability density function (PDF) of the peaks of a random field. It is able to provide a correct estimate of the probability of false detection for the one-, two- and three-dimensional case. We apply this technique to a real two-dimensional interferometric map obtained with ALMA.}

   \titlerunning{A correct computation of the probability of false detection of the matched filter}
   \authorrunning{Vio \& Andreani}
   \maketitle
 
\section{Introduction}

The reliable detection of signals in any observed data is a critical problem common to many subjects \citep{kay98, tuz01, lev08, mac05}. Some specific examples are radar detection \citep{ric05}, wireless communication \citep{zha16}
and particle detection \citep{spi12}. In astronomy, this problem arises when looking for faint sources, for instance in the detection of galaxy clusters \citep{mil10}, X-ray point-sources \citep{ste06}, point-sources in cosmic microwave background \citep{vio04}, extra-solar planets \citep{jen96}, asteroids \citep{gur05} and others. In many practical situations, the technique that is expected to have the best performance, in the sense of providing the greatest probability of true detection for a fixed probability of false detection, is the matched filter (MF). This is the linear filter that maximises the signal-to-noise ratio (SNR) and therefore the detectability of the signal of a known structure embedded in Gaussian random noise.

This technique, however, is based on the assumption that the position of a signal of interest within a sequence of data (e.g. an emission line in a spectrum) or an image (e.g. a point-source in an interferometric map) is known. Often, in practical application this condition is not satisfied. For this reason, the MF is used assuming that, if present, the position of a signal corresponds to a peak of the filtered data. Here we show that, when based on the standard but wrong assumption that the probability density function (PDF) of the peaks of a Gaussian noise process is a Gaussian, this approach may lead to severely underestimating the probability of false detection. The correct method to accurately compute this quantity is also presented.

In Sec.~\ref{sec:MF} the main characteristics of MF are reviewed as well as the reason why it provides an underestimate of the probability of false detection. The alternative method to compute this quantity is detailed  in Sec.~\ref{sec:dan}. Finally, in Sec.~\ref{sec:practical} the procedure is applied to an observed two-dimensional interferometric map obtained with ALMA and the discussion deferred to Sec.~\ref{sec:conclusions}.

\section{Matched filter: an optimal solution of the detection problem} \label{sec:MF}
\subsection{Mathematical formalization}  \label{sec:MF1}

In this section the basic properties of MF are described. For ease of formalism, arguments will be developed in the context of one-dimensional signals. Extension to higher-dimensional situations is formally trivial and will be briefly highlighted in Sec.~\ref{sec:comments}.

The detection problem of a one-dimensional deterministic and discrete signal of known structure $\ssb = [s(0), s(1), \ldots, s(N-1)]^T$,  with length $N$ and the symbol $^T$ denoting a vector or matrix transpose, is based on the following conditions:
\begin{enumerate}
\item The signal of interest has the form $\ssb = a \gb$ with ``$a$'' a positive scalar quantity (amplitude) and $\gb$ typically a smooth function often somehow normalized (e.g., $\max{ \{ g(0), g(1), \ldots, g(N-1) \} } = 1$); \\
\item The signal is embedded within an additive noise $\nb$, i.e.\ the observed signal $\xb$ is given by $\xb = \ssb + \nb$. Without loss of generality, it is assumed that ${\rm E}[\nb] = 0$, where ${\rm E}[.]$ denotes the expectation operator; \\
\item The noise $\nb$ is the realization of a stationary stochastic process with known covariance matrix
\begin{equation} \label{eq:C}
\Cb = {\rm E}[\nb \nb^T]. 
\end{equation}
\end{enumerate}   
Under these conditions, the detection problem consists in deciding whether $\xb$ is pure noise $\nb$ (hypothesis $H_0$) or it also contains a contribution from a signal $\ssb$ (hypothesis $H_1$). In this way, it is equivalent to a decision problem between the two hypotheses:
\begin{equation} \label{eq:decision}
\left\{
\begin{array}{ll}
\Hc_0: & \quad \xb = \nb; \\
\Hc_1: & \quad \xb = \nb + \ssb.
\end{array}
\right.
\end{equation}

\begin{figure*}
        \resizebox{\hsize}{!}{\includegraphics{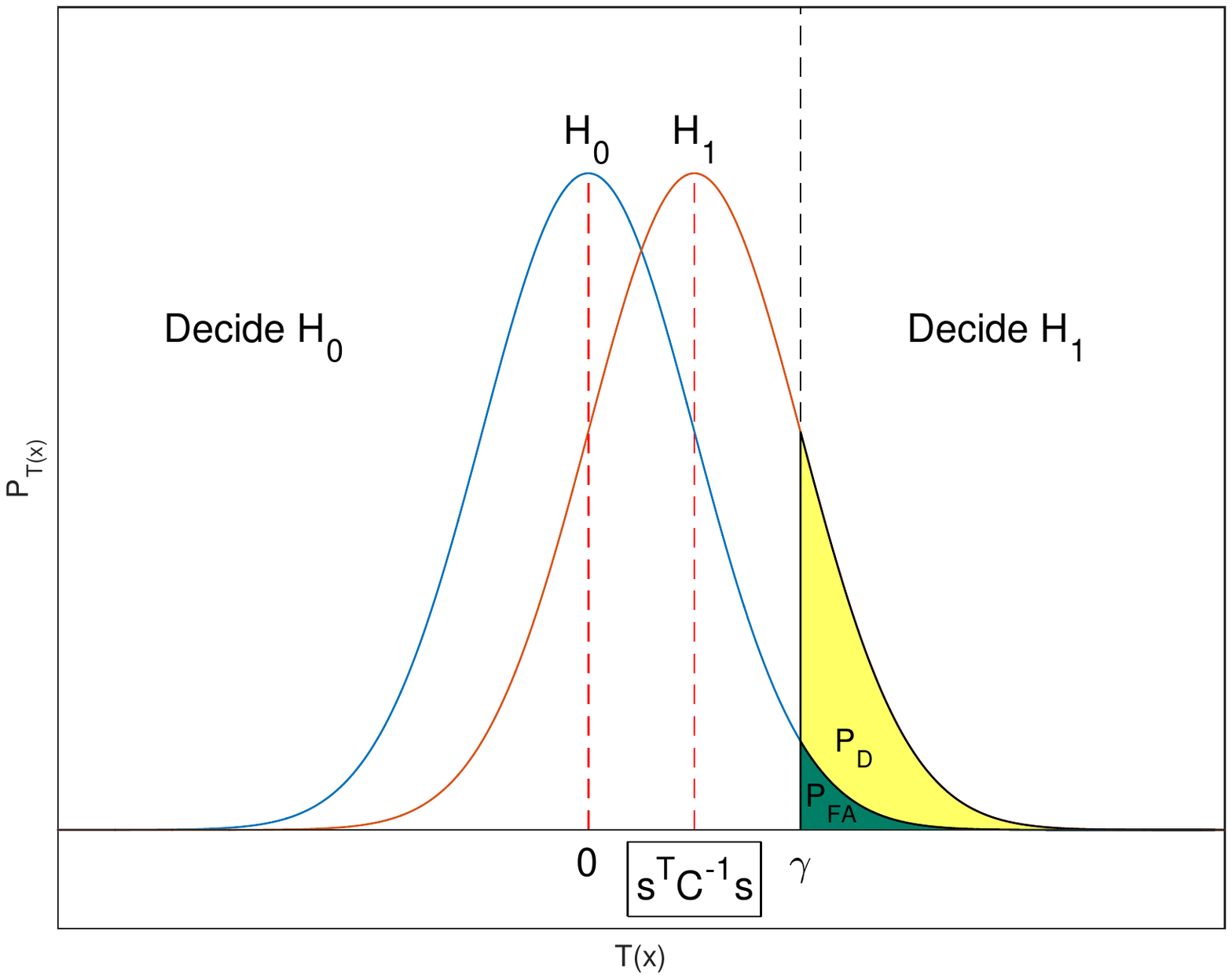}}
        \caption{Probability density function of the statistics $T(\xb)$ under the hypothesis $\Hc_0$ (noise-only hypothesis) and $\Hc_1$ (signal-present hypothesis). 
The detection-threshold is given by $\gamma$. The {\it probability of false alarm} ($\PFA$), called also probability of false detection,
and the {\it probability of detection} ($\PD$) are shown in green and yellow colors, respectively.}
        \label{fig:fig_gauss}
    \end{figure*}

Any decision requires the definition of a criterion and, in this case, the Neyman-Pearson criterion is an effective choice. It consists in the maximization of the {\it probability of detection} $\PD$ under the constraint that the {\it probability of false alarm} $\PFA$ (i.e., the probability of a false detection) does not exceed a fixed value $\alpha$. According to the Neyman-Pearson theorem \citep[e.g., see ][]{kay98}, if $\nb$ is a Gaussian process with covariance function $\Cb$, $\Hc_1$ has to be chosen when
\begin{equation} \label{eq:test1}
\Tc(\xb) = \xb^T \fb > \gamma,
\end{equation}
with
\begin{equation} \label{eq:mf}
\fb = \Cb^{-1} \ssb.
\end{equation}
Here, $\fb$ is the matched filter. The detection threshold $\gamma$ for a fixed $\PFA = \alpha$ is given by
\begin{equation} \label{eq:gammap}
\gamma = \Phi^{-1}(\alpha) \sqrt{ \ssb^T \Cb^{-1} \ssb },
\end{equation}
where $\Phi^{-1}(.)$ is the inverse of the {\it Gaussian complementary cumulative distribution function}
\begin{equation}
\Phi(x) = \int_x^{\infty} \phi(t) dt,
\end{equation}
with 
\begin{equation}
\phi(t) =\frac{1}{\sqrt{2 \pi}} \exp{- \frac{1}{2} t^2}.
\end{equation}
Equation~\eqref{eq:gammap} is due to the fact that the statistic $\Tc(\xb)$ is a Gaussian random variable with variance $\ssb^T \Cb^{-1} \ssb$ and expected value equal to zero under the hypothesis $\Hc_0$, or $\ssb^T \Cb^{-1} \ssb$ under the hypothesis $\Hc_1$ (see Fig.~\ref{fig:fig_gauss}).

When the threshold $\gamma$ is fixed, the probability of false detection, $\alpha$, can be computed by means of
\begin{equation} \label{eq:pfa}
\alpha = \Phi \left( \frac{\gamma}{\left[ \ssb^T \Cb^{-1} \ssb \right]^{1/2}} \right).
\end{equation}
For $\PFA=\alpha$, the probability of detection $\PD$ is 
\begin{equation} \label{eq:pd}
\PD = \Phi \left( \Phi^{-1} \left( \alpha \right) - \sqrt{ \ssb^T \Cb^{-1} \ssb} \right).
\end{equation}

If one sets 
\begin{equation} \label{eq:MFs}
\fb = [0, 0, \ldots, 0, 1, 0, \ldots, 0, 0]^T,
\end{equation}
where the only value different from zero is that corresponding to the greatest value of $\xb$, the operation \eqref{eq:test1} becomes a simple thresholding test which consists of checking if the maximum of the observed data $\xb$ exceeds a fixed threshold. This simplified version of the MF is adopted in some particular situations (see below).

\subsubsection{Properties of the matched filter} \label{sec:comments}

The main characteristics of the MF which are relevant for our discussion are:
\begin{itemize}
\item  The extension of MF to the two-dimensional signals $\Xmatb$ and $\Smatb$ is conceptually trivial. Indeed, setting\footnote{${\rm VEC}[\Fmatb]$ is the operator that transforms a matrix $\Fmatb$ into a column array by stacking its columns one underneath the other.}
\begin{align}
\ssb & = {\rm VEC}[\Smatb]; \label{eq:stack1} \\
\xb & = {\rm VEC}[\Xmatb]; \label{eq:stack2} \\
\nb & = {\rm VEC}[\Nmatb], \label{eq:stack3}
\end{align}
formally the problem is the same as the one-dimensional case given by Eq.~\eqref{eq:decision}. The only difference is that in the one-dimensional case $\Cb$ is a Toeplitz matrix \citep{ant06} whereas in the two-dimensional case it becomes a block Toeplitz with Toeplitz blocks \citep{ram11} that are difficult to work with. The situation rapidly worsens for higher dimensional cases. Because of this, for problems of dimensionality higher than one, it is preferable to work in the Fourier domain \citep{kay98}; \\
\item $\Tc(\xb)$ is a {\it sufficient statistic} \citep{kay98}. Loosely speaking, this means that $\Tc(\xb)$ is able to summarise all the relevant information in the data concerning the decision described in Equation~(\ref{eq:decision}). No other statistic can perform better; \\
\item If the amplitude ``$a$'' of the signal is unknown, then the test in Eq.~(\ref{eq:test1}) can be rewritten in the form
\begin{equation} \label{eq:test2}
\Tc(\xb) = \xb^T \fb > \gamma',
\end{equation}
where now the MF given by Eq.~\eqref{eq:mf} becomes
\begin{equation} \label{eq:mf2}
\fb = \Cb^{-1} \gb,
\end{equation}
and
\begin{equation} \label{eq:gamma3}
\gamma' = \gamma/ a = \Phi^{-1}(\alpha) \sqrt{ \gb^T \Cb^{-1} \gb }.
\end{equation}
In other words, a statistic independent of ``$a$'' is obtained. For the Neyman-Person theorem, in the case of unknown amplitude of the signal, $\Tc(\xb)$ still maximises $\PD$ for a fixed $\PFA$. The only consequence is that $\PD$ cannot be evaluated in advance. In principle this can be done a posteriori by using the maximum likelihood estimate of the amplitude, 
$\widehat{a} = \xb^T \Cb^{-1} \gb / \gb^T \Cb^{-1} \gb$; \\
\item In the derivation of Eq.~\eqref{eq:mf} it has been assumed that both $\xb$ and $\ssb$ have the same length $N$. This implicitly means that the position of the signal $\ssb$  within the data sequence $\xb$ is known.
\end{itemize}
Often, the condition in the last point is not satisfied. In the next section we explore the consequences of this fact.

\subsubsection{The application of the matched filter}  \label{sec:experiment} 

In real data, the signal of interest $\ssb$ has a length $M$ smaller than the length $N$ of the observed data $\xb$  (e.g., an emission line in an experimental spectrum). Moreover, often the position of $\ssb$ within the sequence $\xb$ as well its amplitude ``$a$'' are unknown. In this case the decision problem~\eqref{eq:decision} needs to be modified to
\begin{equation} \label{eq:decision2}
\left\{
\begin{array}{lll}
\Hc_0: & \quad x(i) = n(i); & i=0, 1, \ldots, N-1; \\
\Hc_1: & \quad x(i) = a  g(i-i_0) + n(i) & i=0, 1, \ldots, N-1,
\end{array}
\right.
\end{equation}
where $g(i)$ is nonzero over the interval [0, M-1] and $i_0$ is the unknown delay. As a consequence, the statistic in \eqref{eq:test1} cannot be applied. The common practice to avoid this problem is based on the following four steps:
\begin{enumerate}
\item Computation of the sequence $\Tc(\xb, i_0)$ through the correlation of $\xb$ with the MF given by~\eqref{eq:mf2}
\begin{equation}
 \Tc(\xb, i_0) = \sum_{i=i_0}^{i_0 + M - 1} x(i) f(i-i_0); \quad i_0 = 0, 1, \ldots, N-M.
\end{equation}
This is a linear filtering operation that modifies the characteristics of $\nb$. For example, if $\nb$ has a white-noise spectrum, after the MF operation it becomes coloured (i.e with a non-flat power spectrum); \\
\item Determination of the values $\hat{i}_0$ that maximise $\Tc(\xb, i_0)$. This operation produces the statistic
\begin{equation}
\Tc(\xb, \hat{i}_0) = \underset{ i_0 \in [0, N-M] }{\max} \Tc(\xb, i_0).
\end{equation}
Typically, $\Tc(\xb, \hat{i}_0)$ corresponds to the value of the highest peak in $\Tc(\xb, i_0)$; \\
\item A detection is claimed if
\begin{equation} \label{eq:test2a}
\Tc(\xb, \hat{i}_0) > \gamma',
\end{equation}
or more commonly, since the quantity $ \gb^T \Cb^{-1} \gb$ can be estimated by means of the sample variance $\hat{\sigma}^2_{\Tc}$  of $ \Tc(\xb, i_0) $, if
\begin{equation} \label{eq:test2b}
\Tc(\xb, \hat{i}_0) = \xb^T \fb > u \hat{\sigma}_{\Tc},
\end{equation}
with $\fb$ given by Eq.~\eqref{eq:mf2}, and where $u$ is a value in the range $[3, 5]$; \\
\item The corresponding probability of false detection is computed by means of
\begin{equation} \label{eq:fd1}
\alpha = \Phi \left( \frac{\gamma'}{\left[ \gb^T \Cb^{-1} \gb \right]^{1/2}} \right)
\end{equation}
 if the test in Eq.~\eqref{eq:test2a} is used and 
\begin{equation} \label{eq:fd1}
\alpha = \Phi(u)
\end{equation}
in the other case.
\end{enumerate}
However, the last step is not correct.
This is because with the test~\eqref{eq:test2} one is checking if at the true position of the hypothetical signal $\ssb$, the statistic $\Tc(\xb)$ exceeds the detection threshold. Under the hypothesis $H_0$ (i.e. no signal is present in $\xb$), there is no reason why such a position must coincide with a peak. Indeed, it corresponds to a generic point of the Gaussian noise process. This is the reason why, as shown in Sec.~\ref{sec:MF1}, the PDF of $\Tc(\xb)$ is a Gaussian. On the other hand, with the test~\eqref{eq:test2a} one checks whether the largest peak of $\Tc(\xb, i_0)$ exceeds the detection threshold. Now, contrary to the previous case, under the hypothesis $H_0$, the position $\hat{i}_0$ does not correspond to a generic point of the Gaussian noise process, but rather to the subset of its local maxima. Since the PDF of the local maxima of a Gaussian random process is not a Gaussian, the PDF of $\Tc(\xb, \hat{i}_0)$ cannot be a Gaussian. In other words, the tests~\eqref{eq:test2} and  \eqref{eq:test2a} are not equivalent. This problem becomes even more critical if the number of signals of interest is unknown since the steps $3$-$4$ have to be applied to all the peaks in $ \Tc(\xb, i_0) $.

\subsubsection{A correct computation of the probability of false detection} \label{sec:dan}

In a recent paper \citet{che15a, che15b} provide the explicit PDF of the values $z$ of the peaks in a $N$-dimensional Gaussian isotropic random field of zero-mean and unit-variance for the case $N=1$, $2$, and $3$. For $N=1$,
\begin{equation} \label{eq:pdf_z1}
\psi(z) = \frac{\sqrt{3 - \kappa^2}}{\sqrt{6 \pi}} {\rm e}^{-\frac{3 z^2}{2(3 - \kappa^2)}} + \frac{2 \kappa z \sqrt{\pi}}{\sqrt{6}} \phi(z) \Phi\left(\frac{\kappa z}{\sqrt{3 - \kappa^2}} \right),
\end{equation}
whereas for $N=2$
\begin{multline} \label{eq:pdf_z2}
\psi(z) = \sqrt{3} \kappa^2 (z^2-1) \phi(z) \Phi \left( \frac{\kappa z}{\sqrt{2 - \kappa^2}} \right) + \frac{\kappa z \sqrt{3 ( 2 - \kappa^2)}}{2 \pi} {\rm e}^{-\frac{z^2}{2 - \kappa^2}}\\
+\frac{\sqrt{6}}{\sqrt{\pi (3 - \kappa^2)}} {\rm{e}^{-\frac{3 z^2}{2 (3-\kappa^2)}}} \Phi\left( \frac{\kappa z}{\sqrt{(3 - \kappa^2) (2 - \kappa^2)}} \right).
\end{multline}
For the case $N=3$, the reader should read the original works. Here,
\begin{equation} \label{eq:kd}
\kappa = - \frac{\rho'(0)}{\sqrt{\rho''(0)}},
\end{equation}
where $\rho'(0)$ and $\rho''(0)$ are, respectively, the first and second derivative with respect to $r^2$ of the two-point correlation function $\rho(r)$ at $r=0$, with $r$ the inter-point distance. The same authors also provide the expected number $N_p$ of peaks per unit area. For $N=1$
\begin{equation} 
E[N_p] = \frac{\sqrt{6}}{2 \pi} \sqrt{- \frac{\rho''(0)}{\rho'(0)}},
\end{equation} 
whereas for $N=2$
\begin{equation} \label{eq:number}
E[N_p] = -\frac{\rho''(0)}{\pi \sqrt{3} \rho'(0)}.
\end{equation} 
All these equations hold under the condition that $\rho(r)$ be sufficiently smooth and that $\kappa \le 1$ \citep[see][]{che15a, che15b}.

\begin{figure*}
        \resizebox{\hsize}{!}{\includegraphics{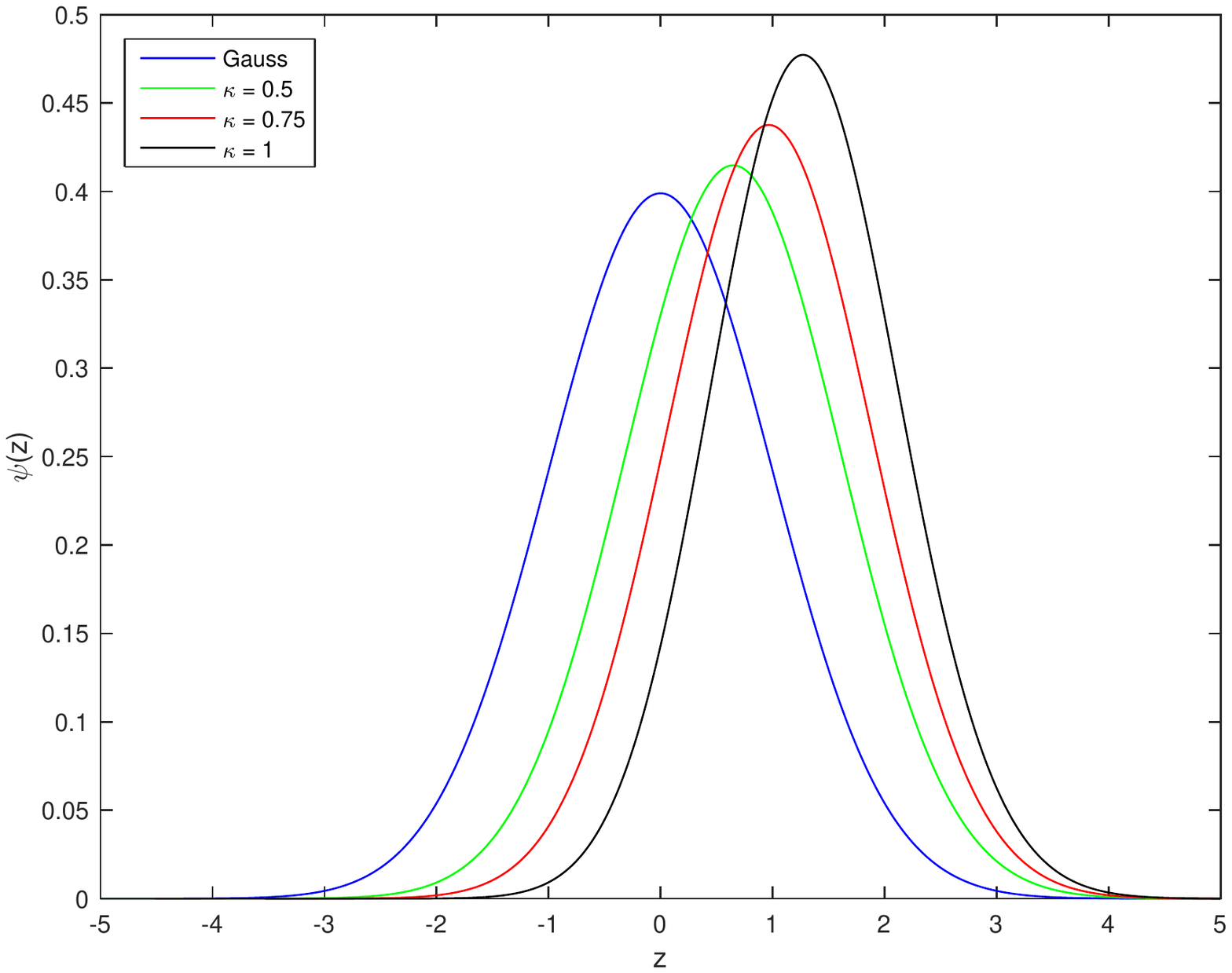}}
        \caption{Comparison of the standard Gaussian  PDF with those of the peaks of an isotropic two dimensional zero-mean and unit-variance Gaussian random field when  $\kappa = 0.5$ , $0.75$, and $1$ (see text).}
        \label{fig:gauss_peak}
    \end{figure*}

\begin{figure*}
        \resizebox{\hsize}{!}{\includegraphics{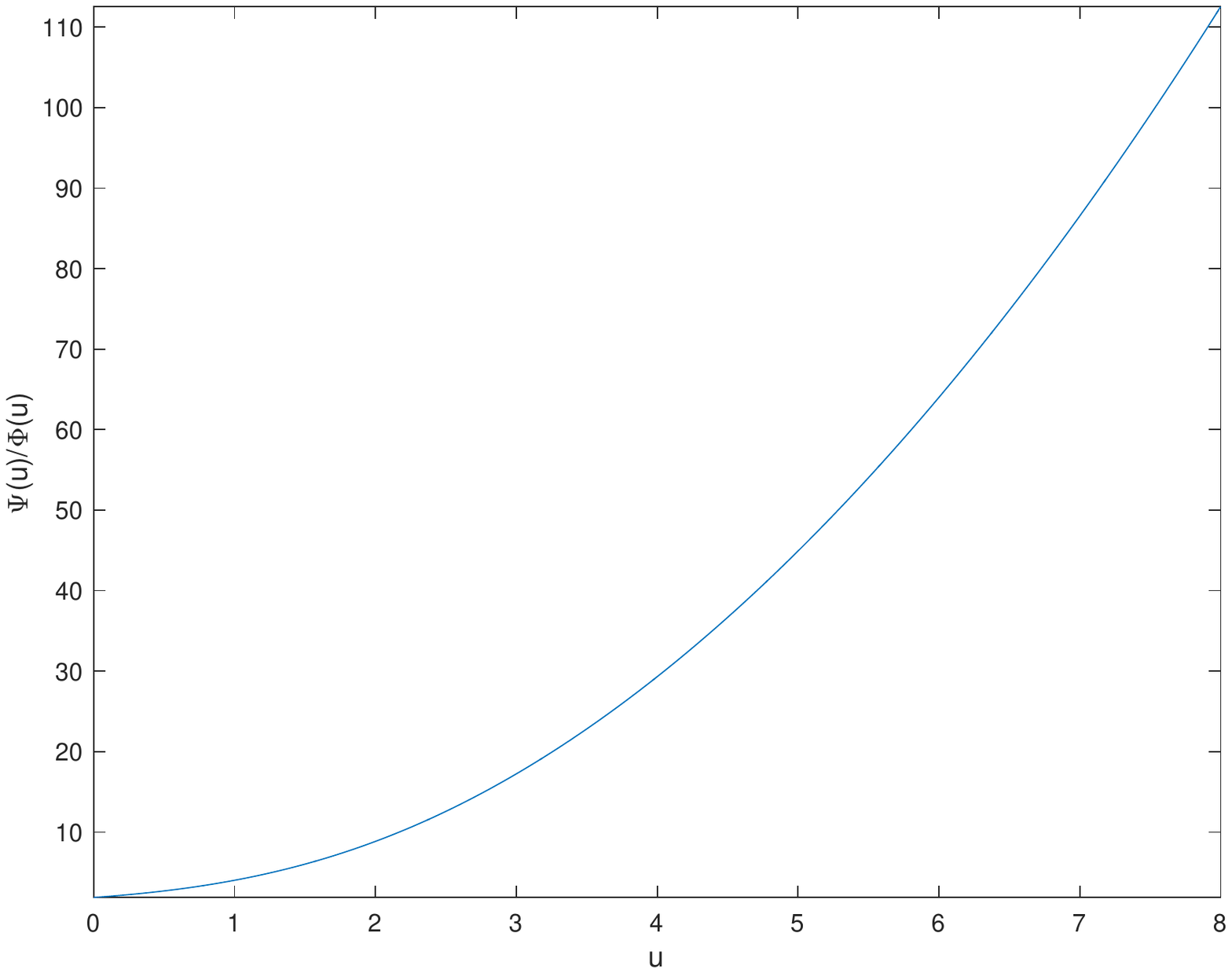}}
        \caption{Ratio $\Psi(u) / \Phi(u)$ of the two probabilities of false detection for the two cases presented in sections~\ref{sec:experiment} and~\ref{sec:dan} as function of the threshold $u$ (see text for detail).}
        \label{fig:roc}
    \end{figure*}

On the basis of these results, the probability that a peak due to a zero-mean unit-variance Gaussian noise process exceeds a fixed threshold ``$u$'' can be computed with
\begin{equation}
\Psi(u)=\int_u^{\infty} \psi(z) dz.
\end{equation} 
Hence, the fourth step in the above procedure needs to be substituted with: 
\begin{itemize}
\item[4.] The corresponding probability of false detection is 
\begin{equation}
 \alpha = \Psi(u).
\end{equation}
\end{itemize}
Figure~\ref{fig:gauss_peak} compares the PDFs $\psi(z)$ for the case $N=2$ and $\kappa=0.5$, $0.75$, and $1$, with the standard Gaussian one. From this figure, the risk of severely overestimating the reliability of a detection is evident. This is supported also by Fig.~\ref{fig:roc} that shows the ratio $\Psi(u) / \Phi(u)$ as a function of the threshold $u$. For instance, for $u = 4$, the probability of false detection provided by $\Phi(u)$ is about $30$ times smaller than that of $\Psi(u)$.

The main problem in the application of this procedure is the computation of the quantity $\kappa$ that in turn requires the knowledge of the analytical form of $\rho(r)$. If the original noise $\nb$ can be written as $\nb = \Hb \wb$, with $\wb$ a white-noise process and $\Hb$ a matrix that implements the discrete form of a known linear filter $h(r)$, then $\rho(r) = \left[h(r) \otimes f(r) \right] \otimes \left[ h(r) \otimes f(r) \right]$ where ''$\otimes$'' represents the correlation operator, and $f(r)$ the continuous form of the MF (e.g., the theoretical point spread function of the instrument). An alternative method, unavoidable if $h(r)$ is unknown, is to fit the discrete sample two-point correlation function of $\Tc(\xb, i_0)$ with an appropriate analytical function. In any case, we have to stress that, as seen above, the knowledge or the estimation of the correlation function of the noise is required also by the MF and therefore is not an additional condition of the procedure.

Strictly speaking, the equations above apply only to continuous random fields. However, it can be reasonably expected that they can also be applied with good results to the discrete random fields if $\rho(r)$ is not too ``narrow'' with respect to the pixel size, or too ``wide'' with respect to the area spanned by the data. In other words, the correlation length of the random field must be greater than the pixel size and smaller than the data extension. Numerical experiments show that good results are obtainable also when the correlation length is comparable to the pixel size.

If the correlation length is smaller than the pixel size, the resulting random field consists of a discrete white noise and the above equations cannot be applied. Since in a discrete random field there is a peak where the value of a pixel is the greatest among the adjacent ones, the corresponding $\psi(z)$ can be computed by means of the order statistics \citep{hog13}. For example, in the two-dimensional case
\begin{equation} \label{eq:discrete}
\psi(z)= 9 \left[ \Phi(z) \right]^8 \phi(z)
\end{equation}
is the PDF of the largest value among nine independent realizations of a zero-mean unit-variance Gaussian process. 
 
\section{Application to an ALMA observation} \label{sec:practical} 

\begin{landscape}
\begin{figure}
        \resizebox{\hsize}{!}{\includegraphics{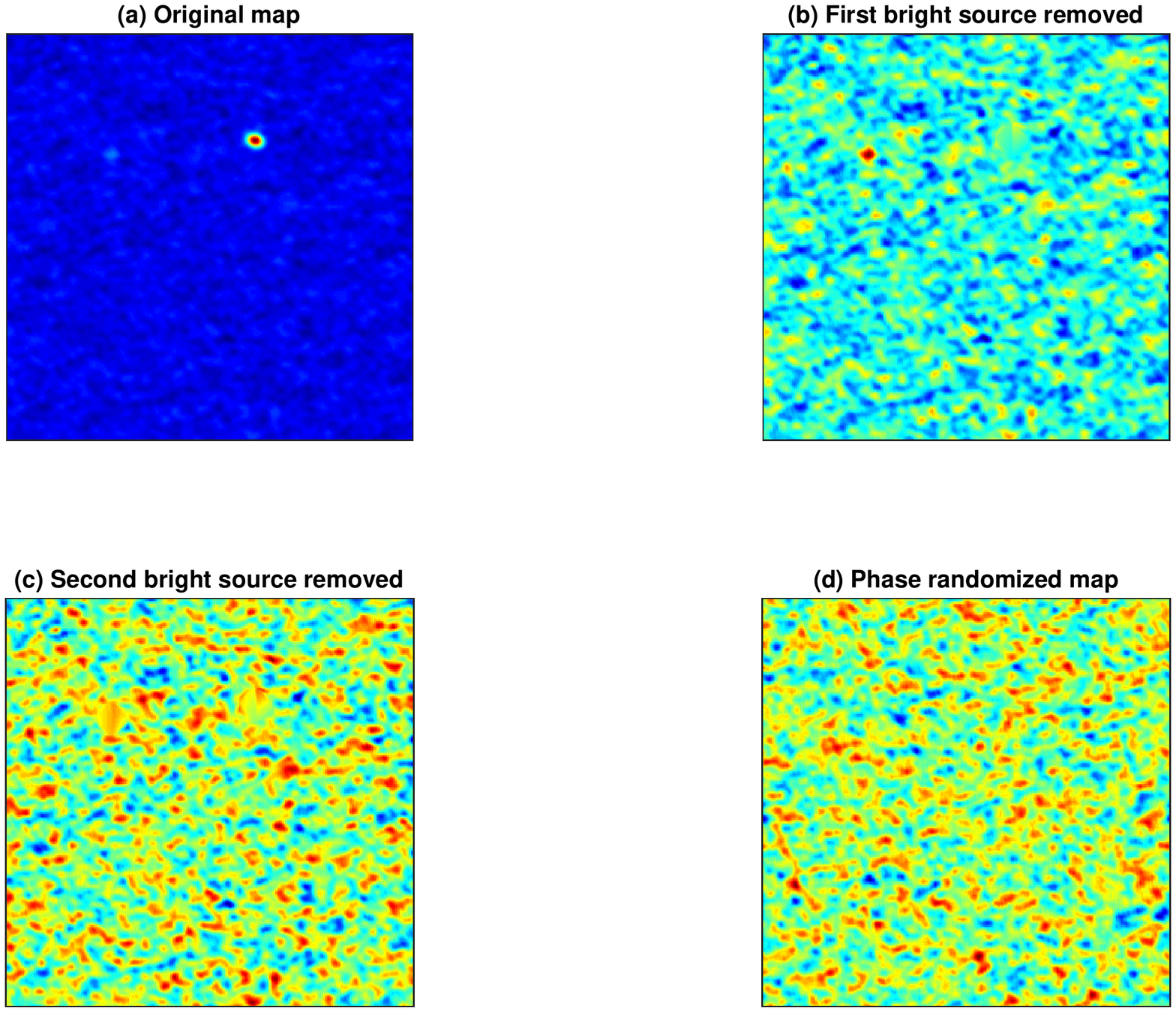}}
        \caption{Panel a): original ALMA map; b) original ALMA map with the brightest source removed; c) original ALMA map, standardized to zero-mean and unit-variance, with both the brightest sources removed; d) phase randomised map (see text).}
        \label{fig:maps_mai}
\end{figure}
\end{landscape}

We apply the above procedure to extract faint (point) sources from a deep map taken with ALMA in Band 6 targeting the Ly-$\alpha$ emitter BDF-3299 \citep{car15, mai15}, which is shown in Figure~\ref{fig:maps_mai}(a) as a $256 \times 256$ pixel map. The total on-source integration time was roughly $300$ minutes and the reached rms value is $7.8 \mu$ Jy/beam. The map has been not corrected for the primary beam, and therefore the resulting noise is uniform across the observed area. We are analysing here only the central part of the map shown in \citet{mai15} and so does not cover the entire area investigated by these authors.

\begin{figure*}
        \resizebox{\hsize}{!}{\includegraphics{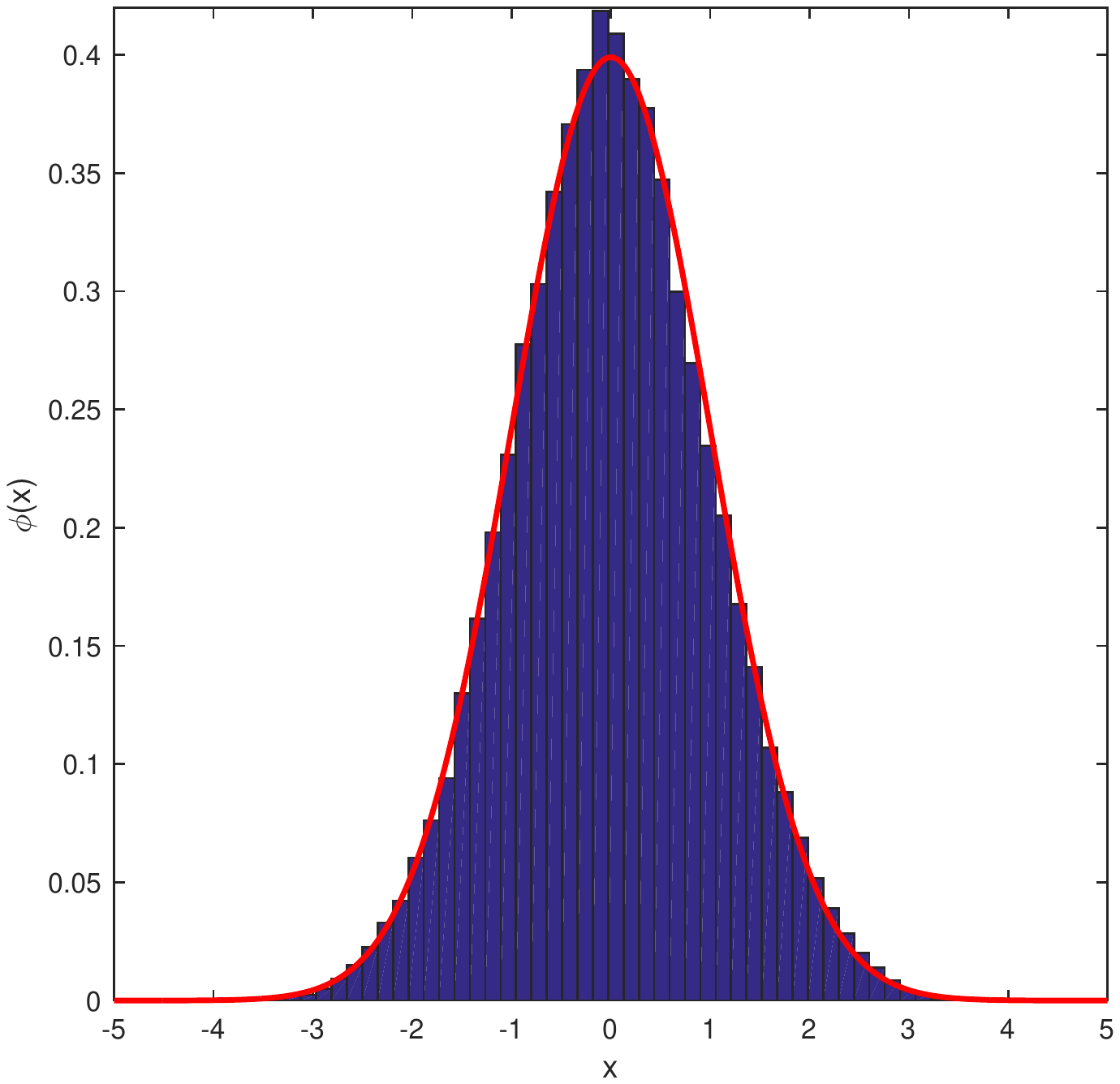}}
        \caption{Histogram of the pixel values of the map in Fig.~\ref{fig:maps_mai}(c) normalised to zero-mean and unit-variance. The red line represents the standard Gaussian probability density function.}
        \label{fig:hist_map_mai}
\end{figure*}

A bright source is apparent and, as shown in  Fig.~\ref{fig:maps_mai}(b), when this is removed and the corresponding area filled with an interpolating two-dimensional cubic spline, another bright source becomes visible. If this is also removed, no additional sources are obvious. In fact, the resulting map in Fig.~\ref{fig:maps_mai}(c), standardized to zero-mean and unit-variance, resembles a Gaussian random field. This impression is confirmed by Fig.~\ref{fig:hist_map_mai}, where the histogram of the pixel values is compared to the standard Gaussian probability density function, as well as by the similarity with Fig.~\ref{fig:maps_mai}(d) which shows a Gaussian random field, obtained by means of the phase-randomization technique\footnote{The phase-randomization technique consists in inverting the Fourier spectrum of a map after the substitution of the discrete Fourier phases with uniform random variates in the range $[0, 2 \pi]$ \citep{pro94}.}, which has the same two-dimensional spectrum of the map in Fig.~\ref{fig:maps_mai}(c).

\begin{landscape}
\begin{figure}
        \resizebox{\hsize}{!}{\includegraphics{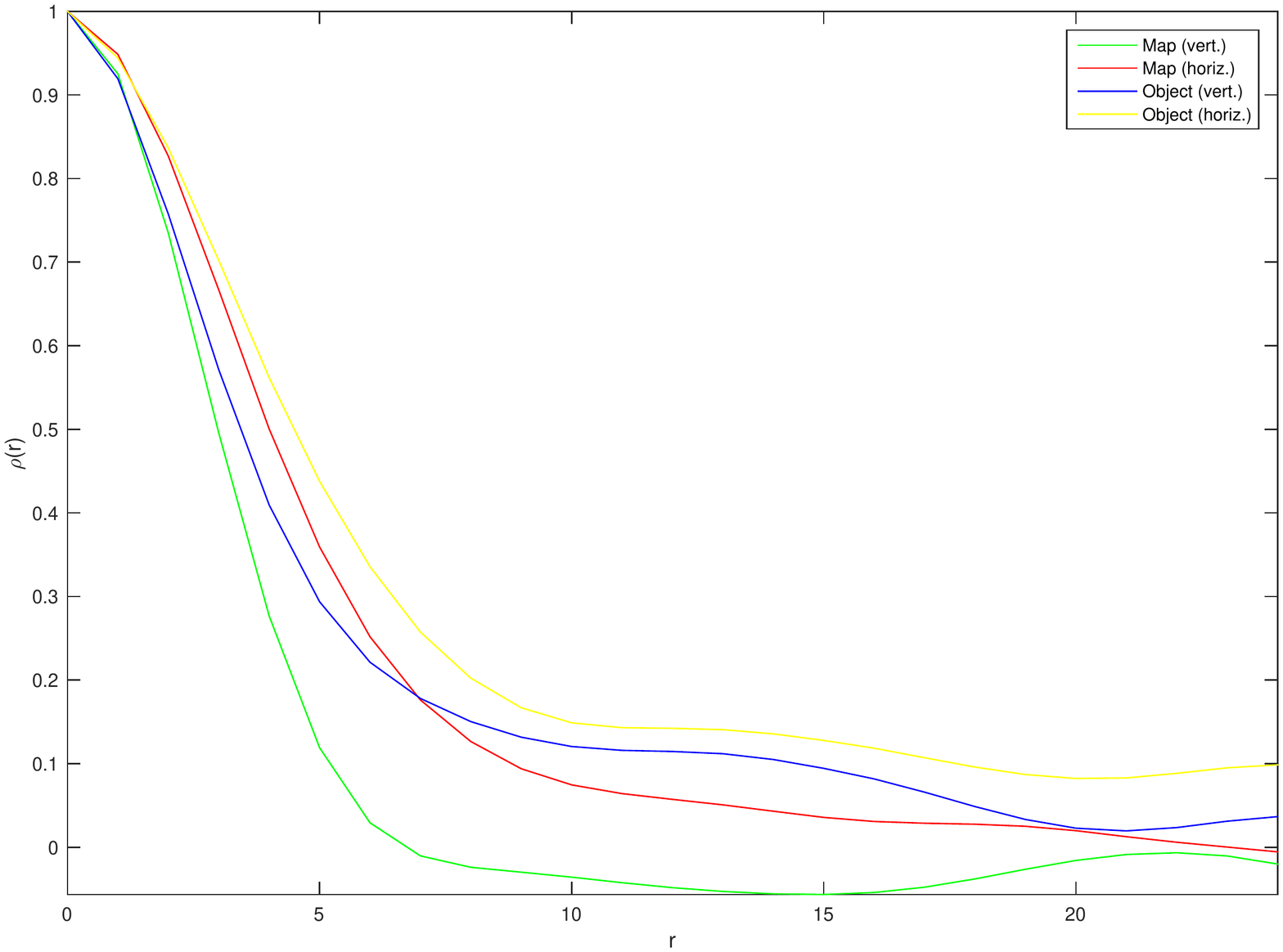}}
        \caption{Autocorrelation function along the vertical and the horizontal directions for both the original ALMA map in Fig.~\ref{fig:maps_mai}(c) and the brightest source visible in Fig.~\ref{fig:maps_mai}(a).}
        \label{fig:autocorr_mai1}
\end{figure}
\end{landscape}

Most of the structures visible in Fig.~\ref{fig:maps_mai}(c) are certainly not due to physical emission. This means that the question we are faced with is the detection of point-sources in Gaussian noise which, however, is not white. As seen in Sec.~\ref{sec:MF}, filtering a Gaussian random field containing a deterministic signal with a MF allows a reduction of the contamination by the unwanted random component and an enhancement of the desired one. In the present case, however, the use of MF is difficult. This is because MF works in such a way as to filter out the Fourier frequencies where the noise is predominant, preserving those where the signal of interest gives a greater contribution. However, as shown by Fig.~\ref{fig:autocorr_mai1}, the autocorrelation functions (ACF) along the vertical and the horizontal directions of the brightest object is similar to those of the underlying random field. This means that the point-sources and the ``blob structures'' due to the  noise have similar shapes. In other words, there is nothing to filter out. For this reason, in the first step in the procedure of Sec.~\eqref{sec:experiment} the MF $\fb$ takes the form~\eqref{eq:MFs} and $\Tc(\xb, i_0) = \xb$. Hence, the detection test becomes a thresholding test where a peak in the map is claimed to be a point-source if it exceeds a given threshold.

\begin{landscape}
\begin{figure}
        \resizebox{\hsize}{!}{\includegraphics{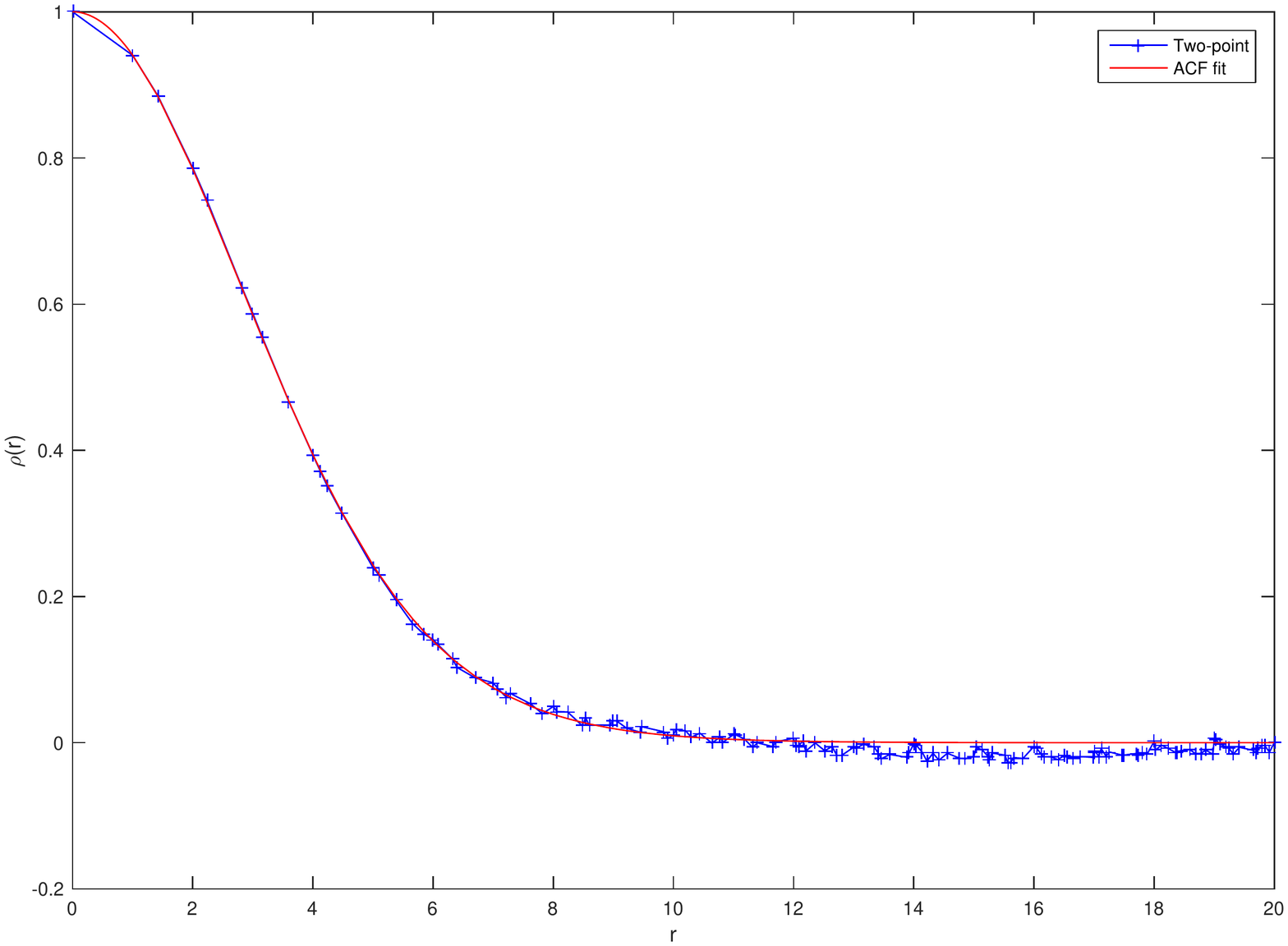}}
        \caption{Sample two-point  correlation functions for the map in Fig.~\ref{fig:maps_mai}(c) vs. the fitted one given by Eq.~\eqref{eq:ACFs}.}
        \label{fig:autocorr_mai2}
\end{figure}
\end{landscape}

\begin{figure*}
        \resizebox{\hsize}{!}{\includegraphics{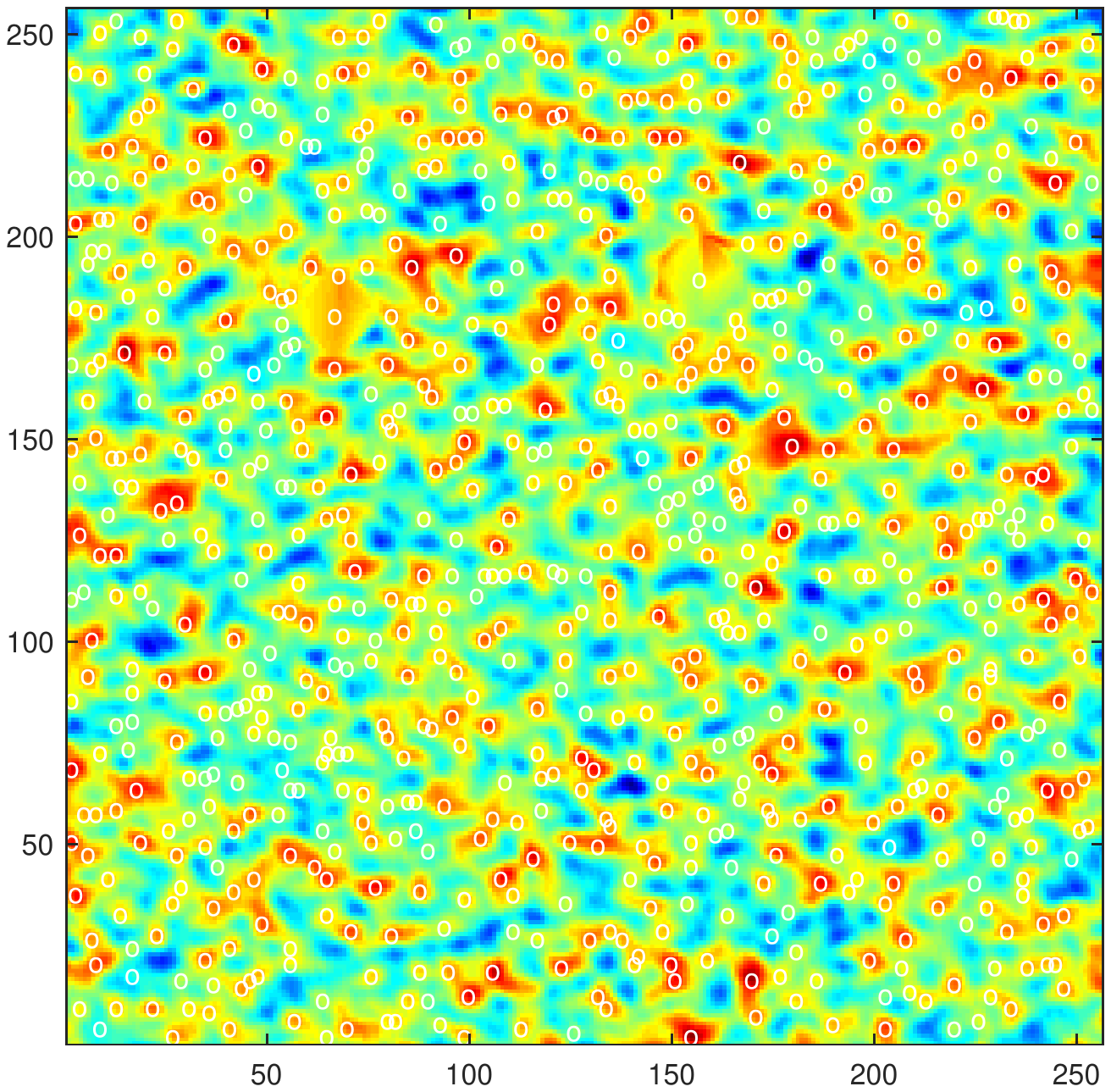}}
        \caption{Map of the ALMA Band 6 observations after the removal of the two brightest sources, as shown in Fig.~\ref{fig:maps_mai}(c).
       The small open circles correspond to the identified peaks.} 
        \label{fig:peak_original_mai}
\end{figure*}

\begin{figure*}
        \resizebox{\hsize}{!}{\includegraphics{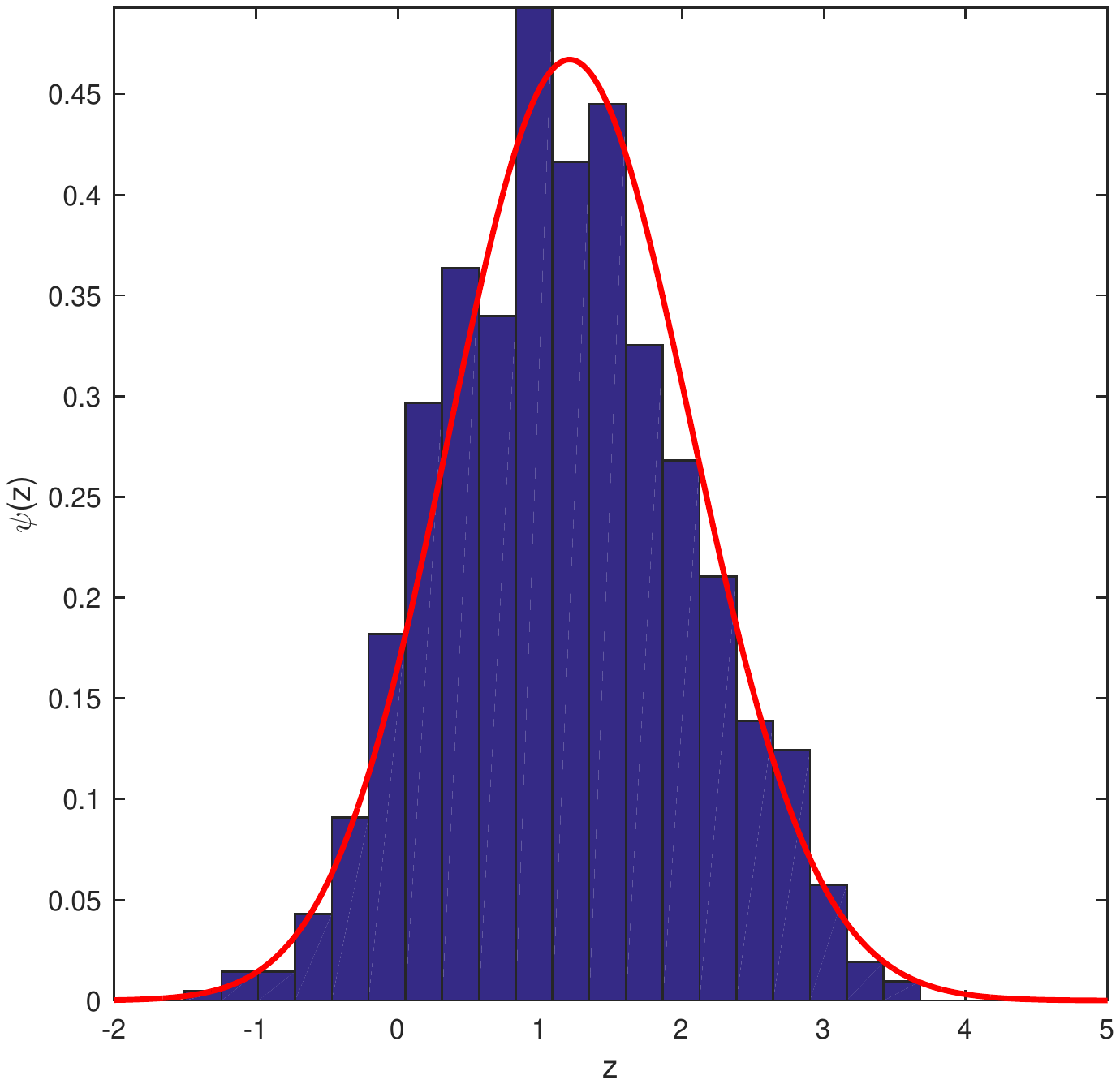}}
        \caption{Histograms of the peak values of the maps in  Fig.~\ref{fig:peak_original_mai}, standardized to zero mean and unit variance,  vs. the theoretical PDF given by Eq.~(\ref{eq:pdf_z2}).}
        \label{fig:peak_pdf_mai}
\end{figure*}

The procedure presented in Sec.~\ref{sec:dan} requires the isotropy of the noise field. As shown in Fig.~\eqref{fig:autocorr_mai1} this condition is approximately satisfied. The small differences between the ACFs along the vertical and horizontal directions are probably due to the fact that, as standard procedure for all the interferometric images, the ALMA map in Fig.~\ref{fig:maps_mai}(a) is the result of a deconvolution \citep[e.g., see ][]{tho04}. As well known, this is a problematic operation. Figure~\eqref{fig:autocorr_mai2} shows that the correlation model
\begin{equation} \label{eq:ACFs}
\rho(r)=b^{-\ln{( 1 + c r^2)}},
\end{equation}
where $b$ and $c$ are free parameters, is able to provide a very good fit to the two-point correlation function of the map. For this model it is
\begin{equation}
\kappa=\frac{\ln{ (b)}}{\sqrt{ \ln{ (b)} + \ln^2{ (b)}}}
\end{equation}
and in the present case it results in $\kappa= 0.95$. The corresponding PDF for the peaks marked in Fig.~\ref{fig:peak_original_mai} is shown in Fig.~\ref{fig:peak_pdf_mai}. Its agreement with the histogram of the peak values is good. 

The conclusion is that it is not possible to claim the presence of point-sources in addition to the two bright ones detected beforehand, because the values of the peaks are compatible with the random fluctuations of a noise field. Indeed, using Eq.~\eqref{eq:number}, the expected number of peaks in the map corresponding to the correlation model~\eqref{eq:ACFs} is given by
\begin{equation}
E[N_p] = \frac{c + \ln{b}}{\pi \sqrt{3}}
\end{equation}
multiplied by the number of pixels. The result is $822$ whereas the number effectively observed is $806$. Since the distribution of peaks in the map appears rather regular, this value is well within the $\pm \sqrt{N_p}$ interval that is the estimate of the standard deviation for the expected number of points generated by a uniform spatial process\footnote{This term indicates a spatial process that produces a regular distribution of points over an area.}. Moreover, the value of the highest peak is $z_{\rm max} = 3.68$ and since $\Psi(z_{\rm max}) \approx 2.62 \times 10^{-3}$, the expected number of peaks that are equal or randomly exceed $z_{\rm max}$ is $2.62 \times 10^{-3} \times 806 \approx 2$, namely a value compatible with that effectively observed. On the other hand if, following the standard procedure, $\Phi(z_{\rm max})=1.17 \times 10^{-4}$ had been used, that number would be $1.17 \times 10^{-4} \times 806 \approx 10^{-1}$. In other words, the peak corresponding to $z_{\rm max}$ should have been considered a detected point-source with a confidence level of $99.99\%$.

As final comment, it is important to stress that these results do not mean that in the ALMA map there are only two point-sources, but only that it is not possible to claim the presence of others at a reliable confidence level.

\section{Conclusions} \label{sec:conclusions}

In this paper we show that, when the position and number of the searched signals/sources within observed data are unknown, the commonly-adopted matched filter may severely underestimate the probability of false detection
if applied in its standard form. As a consequence, statistical significance can be given to structures that are actually due to the noise. Because of this, an alternative method has been proposed which is able to provide a correct estimate of this quantity. Its application to a map taken by ALMA in Band 6 towards a faint extragalactic source demonstrates the risk of spurious detections when the probability of false detection is incorrectly estimated.

\begin{acknowledgements}
This research has been supported by a ESO DGDF grant 2014 and R.V. thanks ESO for hospitality. The authors thank Stefano Carniani and Roberto Maiolino for providing the ALMA data before publication and for useful discussions, and Andy Biggs
for careful reading of the manuscript. This paper makes use of the following ALMA data: ADS/JAO.ALMA\#2012.1.00719.S. ALMA is a partnership of ESO (representing its member states), NSF (USA) and NINS (Japan), together with NRC (Canada) and NSC and ASIAA (Taiwan) and KASI (Republic of Korea), in cooperation with the Republic of Chile. The Joint ALMA Observatory is operated by ESO, AUI/NRAO and NAOJ.
\end{acknowledgements}

\clearpage
\clearpage
\clearpage


\end{document}